\documentclass[twocolumn, appendixfloats, revtex4, numberedappendix, twocolappendix, iop]{openjournal}

\usepackage{CJK}

\usepackage{enumitem}
\usepackage{url}

\usepackage[most,breakable]{tcolorbox}

\newcommand{\Mgeff}{M_{\rm g,eff}}

\shorttitle{Quantifying Bursty Star Formation in Dwarf Galaxies}
\shortauthors{Ting \& Ji}

\begin{document}
\begin{CJK*}{UTF8}{gbsn}

\title{Quantifying Bursty Star Formation in Dwarf Galaxies \vspace{-1.5cm}}

\author{Yuan-Sen Ting (丁源森)}
\affiliation{Department of Astronomy, The Ohio State University, Columbus, OH 43210, USA}
\affiliation{Center for Cosmology and AstroParticle Physics (CCAPP), The Ohio State University, Columbus, OH 43210, USA}
\affiliation{Research School of Astronomy \& Astrophysics, Australian National University, Cotter Rd., Weston, ACT 2611, Australia}

\author{Alexander P. Ji}
\affiliation{Department of Astronomy \& Astrophysics, University of Chicago, 5640 S Ellis Avenue, Chicago, IL 60637, USA}
\affiliation{Kavli Institute for Cosmological Physics, University of Chicago, Chicago, IL 60637, USA}

\begin{abstract}
Dwarf galaxy star formation histories are theoretically expected to be bursty, potentially leaving distinct imprints on their chemical evolution. We propose that episodic starbursts with quiescent periods longer than $\sim$100 Myr should lead to discontinuous tracks in a dwarf galaxy's [$\alpha$/Fe]-[Fe/H] chemical abundance plane, with metallicity gaps as large as 0.3-0.5 dex at [Fe/H] = -2. This occurs due to continued Fe production by Type Ia supernovae during quiescent periods. We demonstrate that Gaussian mixture models can statistically distinguish discontinuous and continuous tracks based on the Akaike Information Criterion. Applying this method to APOGEE observations of the Sculptor dSph galaxy suggests an episodic star formation history with ${\sim}300$ Myr quiescent periods. While current dwarf galaxy datasets are limited by small spectroscopic sample sizes, future surveys and extremely large telescopes will enable determining large numbers of precise chemical abundances, opening up the investigation of very short timescales in early dwarf galaxy formation. This unprecedentedly high time resolution of dwarf galaxy formation in the early Universe has important implications for understanding both reionization in the early Universe and the episodic star formation cycle of dwarf galaxies.
\end{abstract}

\section{Introduction}

Standard galaxy formation in the $\Lambda$CDM paradigm suggests that dwarf galaxies (stellar masses $\sim 10^3-10^9~{\rm M}_\odot$) inherently have bursty or episodic star formation histories. Due to their low masses, stellar feedback can temporarily quench star formation, causing the instantaneous star formation rate to fluctuate by orders of magnitude over 10-100 Myr timescales \citep[e.g.,][]{Governato2010,Hopkins2014,Chan2015,Dutton2016,Tollet2016,Bose2019,Emami2019}. In the faintest dwarf galaxies, supernova feedback can completely quench star formation for hundreds of Myr \citep[e.g.,][]{BlandHawthorn2015,Wheeler2019,Hirai2024} or even permanently \citep{Gallart2021,Chen2022}. Reionization also appears to force many faint dwarf galaxies to pause star formation for a few Gyr before resuming \citep{Savino2023,McQuinn2023,McQuinn2024}.

Such bursty star formation is crucial for reconciling observed dwarf galaxy central density profiles with $\Lambda$CDM predictions. It is this burstiness that causes initially cuspy \citet{Navarro1996} halos to transform into the observed cored density profiles \citep[e.g.,][]{Pontzen2012,Chan2015,Tollet2016}. Furthermore, JWST observations suggest that dwarf galaxies may over-ionize the IGM, with bursty star formation proposed as a possible solution \citep{Sun2023,Clarke2024,Munoz2024,Simmonds2024}. Burstiness has also been linked to the miniquenching of high-redshift low-mass galaxies \citep{Antwi-Danso2024,Dome2024,Looser2024}.

In principle, bursty star formation could imprint upon a dwarf galaxy's chemical abundance distribution, such as the pattern of [$\alpha$/Fe] vs [Fe/H] of individually resolved stars and the stellar metallicity distribution function. Fe is produced by both core-collapse supernovae (CCSNe) and Type~Ia supernovae (SNeIa), while $\alpha$-elements are produced primarily by CCSNe. A galaxy experiencing a significant quiescent period will cease $\alpha$-element production, while Fe production by SNeIa continues. Consequently, when star formation resumes, there could be discrete jumps in the [Fe/H] distribution of newly formed stars. These jumps would be visible in sufficiently populated chemical abundance distributions. Such features might be detectable in current dwarf galaxy data samples obtained over the past decade \citep[e.g.,][]{Cohen2010,Kirby2011,Starkenburg2013,Hendricks2014,Hill2019,Kirby2019,Kirby2020,Hasselquist2021}, as well as in future surveys \citep[e.g.,][]{Skuladottir2023}. 

The majority of chemical evolution models employ smooth prescriptions for gas and star formation histories (SFH) \citep[e.g.,][]{Cescutti2008,Kirby2011,Pilkington2012,Vincenzo2014,Recchi2015,Ishimaru2015,Yuan2016,Cote2017,Weinberg2017,Matteucci2021,delosReyes2022,Johnson2023,Sandford2024}, with some notable exceptions \citep[e.g.,][]{Hirashita2001,Johnson2020}. Some semi-analytic models, based on dark matter merger trees, allow for episodic gas accretion, but do not study the impact on individual metallicity distributions \citep[e.g.,][]{Lu2011,Benson2012,Lee2014,Benson2016,Qin2018,Manwadkar2022}. Cosmological simulations tracking chemistry naturally include bursty star formation \citep[e.g.,][]{Jeon2017,Wheeler2019,Applebaum2021}, but they are still relatively limited in mass and time resolution for studying the early chemical evolution of dwarf galaxies and cannot directly model individual known galaxies.

To bridge this gap, we propose that bursty star formation should be resolvable through chemical abundances. We implement a simple chemical evolution model that enforces bursty star formation to observe its effects (Section~\ref{sec:chem-model}). The model predicts a multimodal chemical pattern (Section~\ref{sec:motivation}), which we suggest can be modeled as a Gaussian mixture model (Section~\ref{sec:measurement-sims}). We apply this test to the Sculptor dSph and find evidence that it experienced bursty early star formation (Section~\ref{sec:measurement-sculptor}). We conclude by discussing prospects for future abundance surveys and comparisons to high-redshift observations (Sections~\ref{sec:discussion} and \ref{sec:conclusion}).

\section{Simple Chemical Evolution Model}
\label{sec:chem-model}

To understand how bursty star formation can lead to a distinctive discontinuous chemical track, we draw intuition from a simplified setup for the chemical evolution of a dwarf galaxy, which allows us to integrate the chemical track analytically by varying only the star formation history and the underlying mixing gas mass. We will trace the $\alpha$-elements with the Mg abundance, as it is relatively easy to measure in stars and uncontaminated by SNIa yields \citep[e.g.,][]{Kirby2019,Skuladottir2019,Weinberg2019}.
We adopt the standard one-zone, instantaneous mixing approximation, which is likely applicable for dwarf galaxies due to their small sizes \citep[e.g.,][]{Kirby2011,Vincenzo2019,Johnson2023}.

We assume a prompt source of stellar yield from core-collapse supernovae (CCSNe), which produces both Mg and Fe 10 Myr after a starburst:
\begin{equation}
D_{\text{CC}}(t) = \delta\left( \Big(\frac{t}{\text{Myr}}\Big) - 10 \right),
\end{equation}

\noindent
and a delayed source for Type Ia supernovae (SNeIa), which contribute only to Fe. We assume a power law delay time distribution with index -1.1 and a minimum delay time of 100 Myr \citep{Maoz2011}:
\begin{equation}
D_{\text{Ia}}(t) = 
\begin{cases}
0 & \text{for }t < 100\text{ Myr} \\
0.429 \left(\frac{t}{\text{Myr}}\right)^{-1.1} & \text{for }t \geq 100\text{ Myr}
\end{cases}
\label{eq:SNIaDTD}
\end{equation}

\noindent
where $0.429$ is a normalization constant. We assume the core-collapse supernova IMF-averaged yield for Fe and Mg per supernova to be $y_{\text{CC,Fe}} = 0.081$ and $y_{\text{CC,Mg}} = 0.192$ with the rate of $r_{\text{CC}} = 0.01~ {\text{M}}_{\odot}^{-1}$ core-collapse supernovae per solar mass formed, and the Type Ia supernovae yield of Fe to be $y_{\text{Ia,Fe}} = 0.749$ with negligible Mg production, and a supernova rate of $r_{\text{Ia}} = 0.126~{\text{M}}_{\odot}^{-1}$.
Yields (in $M_\odot$) and rates are adopted from \citet{Kirby2011}.

We then make three simplifying assumptions. First, we assume the yields have no metallicity dependence. This is a reasonable assumption for Mg and Fe in metal-poor dwarf galaxies, where the metallicity dependence of IMF-integrated yields is relatively weak \citep[e.g.,][]{Nomoto2013}. Then, the total yield of heavy elements is simply the convolution of the star formation rate, $\dot{M}_*(t)$, multiplied by the yield convolved with the delay time distribution:
\begin{equation}
Y_X(t) = \dot{M}_*(t) * \left(y_{\text{CC},X} r_{\text{CC}}  D_{\text{CC}}(t) + y_{\text{Ia},X} r_{\text{Ia}} D_{\text{Ia}}(t)\right)
\end{equation}
\noindent
where $Y_X(t)$ is the total yield of element $X$ at time $t$. The total mass of an element $X$ that has ever been produced by a galaxy at time $t$ can be expressed as:
\begin{equation}
M_X(t) = \int_0^t Y_X(t') \, {\rm d}t'
\end{equation}

Second, we assume that we can independently specify the star formation rate $\dot M_*(t)$ and an effective gas mass $\Mgeff(t)$, which includes not just the cold interstellar medium, but also the circumgalactic or intergalactic gas mass that is affected by the stellar yields from this galaxy.
The abundance of element $X$ relative to hydrogen at time $t$ can then be calculated as:
\begin{equation}
[X/\text{H}](t) = \log_{10}\left(\frac{M_X(t)}{\mu_X \Mgeff(t)}\right) - \log_{10} \left(\frac{N_X}{N_H}\right)_\odot
\end{equation}

\noindent
where $\mu_X$ is the atomic mass of element $X$, $\log_{10} (N_X/N_H)_\odot$ is the solar abundance ratio of element $X$ to hydrogen (adopted from \citealt{Asplund2009}). We assume $\mu_X = 24$ for Mg and $\mu_X = 56$ for Fe. We note this assumption is mathematically equivalent to standard one-zone chemical evolution models. Any equations describing the inflows, outflows, metal loading, and star formation efficiencies just determine specific functional forms for the star formation rate $\dot M_*(t)$ and effective gas mass $\Mgeff(t)$. We note that $\Mgeff$ is essentially a ``Lagrangian mass'' of gas affected by metals produced in this galaxy, which could include not just a cold interstellar medium but also the circumgalactic or intergalactic gas.

Third, and the biggest assumption here, we take the mixing gas mass to be a constant, i.e. $\dot{M}_{\rm g,eff}(t) = 0$. This is motivated by simplicity, allowing us to isolate the impact of bursty star formation on chemical evolution. This simple assumption is sufficient to match observations of Sculptor (Section~\ref{sec:measurement-sculptor}). It may also be a reasonable assumption: while dwarf galaxies are known to lose 90-99\% of their metals in outflows \citep[e.g.,][]{Kirby2011b}, we can instead use a $10-100{\times}$ larger $\Mgeff$ than would normally be in part of a galaxy's interstellar medium. We will revisit the implications of this assumption in Section~\ref{sec:discussion} and future work.

Our choice of a semi-analytical framework, while simplified compared to hydrodynamical simulations, provides two key advantages: analytical tractability that allows us to directly connect observable features to physical parameters, and the ability to isolate the specific impact of bursty star formation on chemical evolution. This approach complements existing hydrodynamical simulations that naturally include bursty star formation but cannot yet necessarily achieve the mass and time resolution needed for studying the early chemical evolution of individual known dwarf galaxies \citep{Jeon2017,Wheeler2019,Hirai2024}.

\begin{figure*}
\centering
\includegraphics[width=1.0\textwidth]{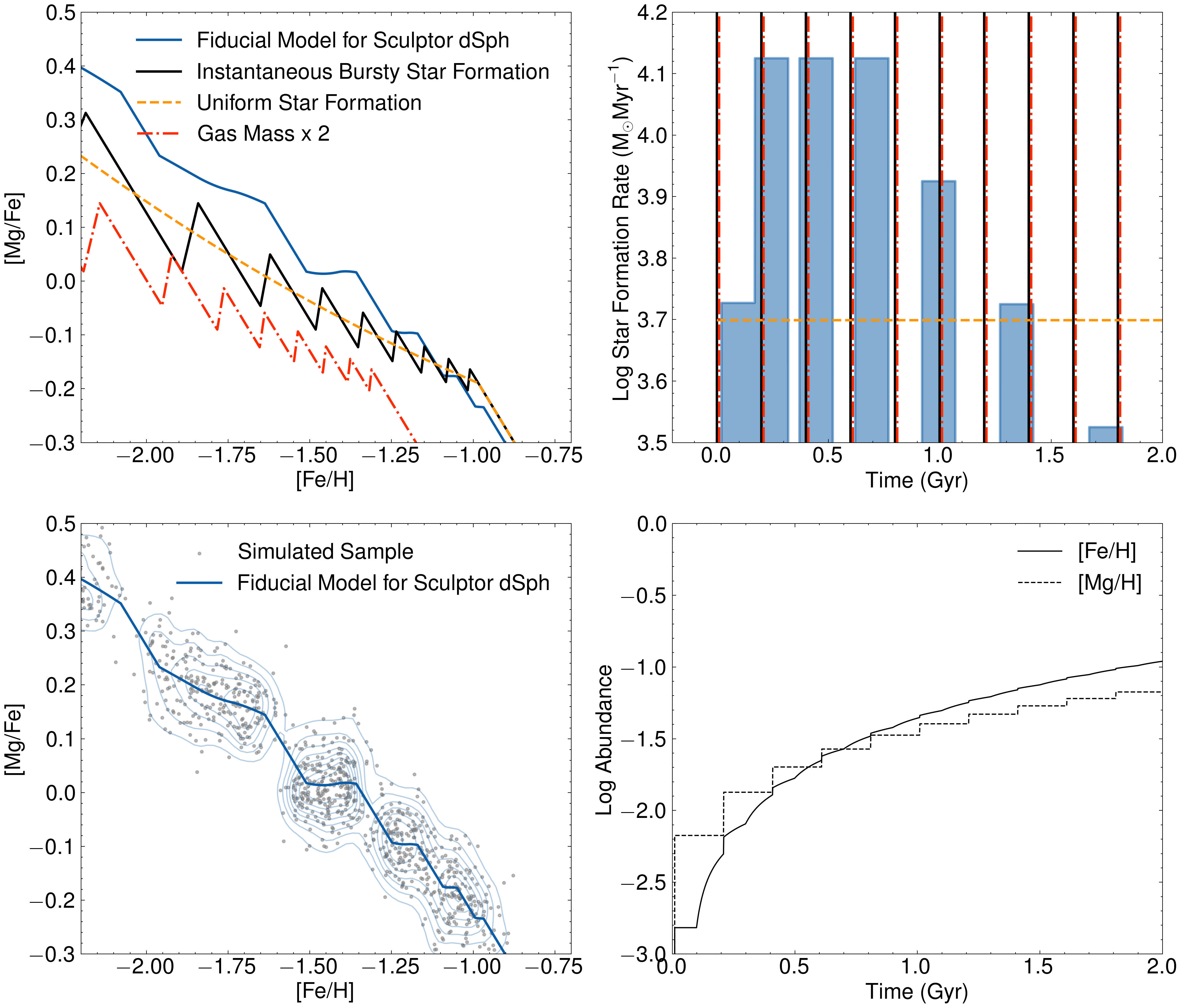}
\caption{Bursty star formation creates discontinuous chemical tracks in the [Mg/Fe]-[Fe/H] plane. The top left panel shows chemical tracks resulting from various star formation histories: our fiducial model to match the Sculptor dSph observation (solid blue line), star formation with instantaneous bursts every 200 Myr over ~2 Gyr (solid black line), continuous star formation (dashed orange line), and a scenario with doubled gas mass (dash-dotted red line). The top right panel displays the corresponding star formation histories. The bottom right panel illustrates the growth of Fe and Mg abundances in the instantaneous starburst case. It demonstrates that the ISM's [Fe/H] continues to accumulate from Type Ia supernovae, while [Mg/H] primarily accumulates from instantaneous core-collapse supernovae. This combination creates gaps in the abundance plane during episodic star formation, where [Fe/H] reflects star formation gaps and [Mg/H] largely reflects yields from individual bursts. The bottom left panel further illustrates this effect, showing 1000 sampled data points (grey) following the fiducial star formation history, with an assumed measurement error of 0.05 dex, along with their kernel density estimation contours. The distinct clumps formed in the [Mg/Fe]-[Fe/H] plane contrast sharply with the continuous star formation case. All modeled star formation rates integrate to a total stellar mass of $10^7 {\rm M}_{\odot}$, reminiscent of the Sculptor dSph galaxy, with an effective gas mass of $3 \times 10^8 {\rm M}_{\odot}$.}
\label{fig1}
\end{figure*}

\section{Expectations for Bursty Star Formation}
\label{sec:motivation}

We now use our simple model to investigate the impact of episodic, bursty star formation on dwarf galaxy chemical evolution. Figure~\ref{fig1} illustrates the impact of bursty star formation on the chemical evolution of a dwarf galaxy using a model with starbursts occurring over a total duration of 2 Gyr. The solid blue line in the top left panel shows the chemical evolution track for our fiducial scenario, where each burst generates approximately $10^6 {\rm M}_{\odot}$ of stars instantaneously, resulting in a total stellar mass of $10^7 {\rm M}_{\odot}$. The model parameters are chosen to resemble a relatively massive dwarf galaxy, similar to the Sculptor dSph, which serves as an archetype for this study. For simplicity, we assume a constant gas mass of $\Mgeff = 3\times 10^8 {\rm M}_{\odot}$ throughout the galaxy's evolution, as this choice produces chemical evolution tracks that are qualitatively similar to the observed data from Sculptor dSph (see Section~\ref{sec:measurement-sculptor})  . The corresponding SFR is visualized in the top right panel. The dotted black line represents a case with more frequent instantaneous starbursts, where we model 10 bursts with a spacing of 200 Myr, each generating $10^6 {\rm M}_{\odot}$ of stars. This scenario sums up to the same total stellar mass as the fiducial model, allowing for a direct comparison of the effects of burst frequency on chemical evolution.

To better understand the mechanisms driving the chemical gaps in an episodic star formation scenario, the bottom right panel of Figure~\ref{fig1} illustrates the growth of Fe and Mg abundances over time in the instantaneous starburst case. This panel demonstrates that the ISM's [Fe/H] continues to accumulate from Type Ia supernovae during quiescent periods, while [Mg/H] primarily accumulates from instantaneous core-collapse supernovae during starbursts. This differential enrichment pattern explains the formation of gaps in the abundance plane during episodic star formation: [Fe/H] effectively traces the star formation gaps, while [Mg/H] reflects individual bursts.

The episodic star formation scenario, as shown in the top panels of Figure~\ref{fig1}, leads to a discontinuous chemical evolution track, manifesting as distinct clumps in the [Mg/Fe]-[Fe/H] plane. This contrasts with the case of uniform star formation, depicted by the orange dashed line, which results in a smooth and continuous chemical evolution track.

Our fiducial chemical evolution model is shown by a solid blue line and will be compared with the Sculptor observations in the following section. The fiducial model differs from the instantaneous model in two key aspects. First, instead of assuming an instantaneous starburst, we consider a more realistic scenario where each starburst lasts for 150 Myr. This choice is motivated by simulations \citep[e.g.,][]{Wheeler2019,Patel2022,Hopkins2023} that have shown that multiple starbursts often occur within a ``block'' of star formation activity, followed by a quiescent period lasting a few hundred million years. Such quiescent periods may also be observed in higher mass dwarf galaxies by JWST \citep[e.g.,][]{Looser2024}. Second, we allow the timing and normalization of the star formation bursts to vary in order to match the star formation history and observed chemical evolution of Sculptor.  We assume a gradual increase in the integrated star formation mass from $10^6 {\text M}_{\odot}$ to $2 \times 10^6 {\text M}_{\odot}$ at 1 Gyr, followed by a decrease to $0.5 \times 10^6 {\text M}_{\odot}$ for the final burst. This results in a total stellar mass of $10^7 {\text M}_{\odot}$ (see top right panel of Figure~\ref{fig1}), which matches the shape of the metallicity distribution function (MDF) and star formation history observed in the Sculptor dSph \citep{Kirby2011,Weisz2014}.

It is important to note that the temporal ``resolution'' of the starburst in our model is limited by the minimum delay time for Type Ia supernovae, which is 100 Myr in this study. Consequently, our model cannot resolve bursts separated by less than this timescale. For the purpose of this analysis, we define two distinct starbursts as events separated by more than 150 Myr, with star formation proceeding uniformly throughout each 150 Myr burst.
In our fiducial model, we assume a total of 7 starburst blocks, with the temporal gap between each block increasing from 150 Myr to 450 Myr in increments of 50 Myr. We find that this increasing temporal gap over the 2 Gyr of star formation in Sculptor leads to a better match with the observed chemical properties of the galaxy. The bottom left panel shows the chemical track of the fiducial case, and overlaid on the track are 1000 sample points drawn from the chemical evolution model, following the star formation rate and assuming an abundance uncertainty $\sigma$ of 0.05 dex for both [Fe/H] and [Mg/Fe], a typical precision (though not necessarily accuracy) from high-resolution spectroscopic surveys like APOGEE. The bursty star formation clearly results in a multimodal distribution of stellar abundances.

As shown in the top left panel, in the instantaneous burst case, the chemical evolution track exhibits sharp features, where the slope of the upward spike in [Mg/Fe] is solely determined by the burst. In contrast, our fiducial model with star formation spanning 150 Myr per starburst allows for continuous enrichment of the ISM by Type Ia supernovae during the burst, resulting in more subdued and concave stripes instead of discrete spikes. However, it is important to note that while the instantaneous starburst model might appear to lead to a marginally larger spread in the chemical track, the track represents the chemistry of the gas, not necessarily the observed stars. In the instantaneous case, all stars are created at the base of each spike, rather than being gradually formed as the burst continues over a more prolonged period. As a result, the fiducial model shown actually has a larger spread in [Mg/Fe] for a given clump compared to the instantaneous case. Finally, when the gas mass is doubled, the gaps between the clumps in the [Mg/Fe]-[Fe/H] plane become less pronounced. This is because, for a fixed iron production rate, the magnitude of the [Fe/H] gap is modulated by the underlying gas mass.

\subsection{Quantifying the gap in [Fe/H] between starbursts}
\label{subsec:feh-gap}

To better understand the impact of bursty star formation on the chemical evolution of dwarf galaxies, we analytically estimate the difference in [Fe/H] between two consecutive starbursts. This difference manifests as the spacing between the peaks of a multimodal MDF in the case of bursty star formation. The spacing between the distinct star formation episodes is determined by the degree to which the metallicity evolves during the quiescent periods separating the starbursts. This evolution is primarily driven by the accumulated reservoir of Type Ia supernovae from previous star formation episodes, which continuously enrich the interstellar medium with iron-peak elements (bottom right panel of Fig~\ref{fig1}).

Consider two starbursts separated by a time gap $\delta t_{\text{gap}}$. The corresponding gap in [Fe/H] between the first and second starburst, to first approximation, is contributed by the supernovae from the first starburst and can be expressed as
\begin{equation}
\Delta\text{[Fe/H]} = \log_{10} \frac{M_{\text{Fe}}(t) + \Delta M_{\text{Fe}}}{M_{\text{Fe}}(t)} \approx \frac{\Delta M_{\text{Fe}}}{M_{\text{Fe}}(t) \ln(10)}
\label{eq:taylor-expansion}
\end{equation}

\noindent
where $\Delta M_{\text{Fe}}$ is the additional iron mass generated between times $t$ and $t + \delta t_{\text{gap}}$. The approximation is a Taylor expansion assuming that the change in total Fe mass during the gap is small compared to the existing Fe mass.

If we assume a starburst generating a total mass $M_*$, and a power-law delay time distribution given by Equation~\ref{eq:SNIaDTD}, the gap in [Fe/H] between two starbursts contributed by type Ia supernovae can be approximated as:
\begin{equation}
\begin{alignedat}{2}
\Delta\text{[Fe/H]}_{\text{Ia}} \approx{} &\frac{y_{\text{Ia,Fe}} r_{\text{Ia}} M_*}{\mu_{\text{Fe}} \Mgeff 10^{\text{[Fe/H]} - 4.5} \ln(10)} \times \\
& \qquad \int_{t_{\text{min}}}^{\delta t_{\text{gap}}} 0.429 \left(\frac{t}{\text{Myr}}\right)^{-1.1} dt
\end{alignedat}
\end{equation}

\noindent
where $\log_{10}(N_{\text{Fe}}/N_{\text{H}})_\odot = -4.5$ so we can express $M_{\text{Fe}}(t)$ in terms of [Fe/H]:
\begin{equation}
M_{\text{Fe}}(t) = \mu_{\text{Fe}} \Mgeff 10^{\text{[Fe/H]} - 4.5}
\end{equation}

\noindent
Evaluating the integral yields:
\begin{equation}
\begin{alignedat}{2}
\Delta\text{[Fe/H]}_{\text{Ia}} \approx & \frac{4.29 y_{\text{Ia,Fe}} r_{\text{Ia}} M_*}{\mu_{\text{Fe}} \Mgeff 10^{\text{[Fe/H]} - 4.5} \ln(10)} \\ & \qquad \left[ \left(\frac{t_{\text{min}}}{\text{Myr}}\right)^{-0.1}-\left(\frac{\delta t_{\text{gap}}}{\text{Myr}}\right)^{-0.1} \right]
\end{alignedat}
\label{eq:deltafeh}
\end{equation}

\noindent
Furthermore, the Fe contribution from CCSNe is 
\begin{equation}
\Delta\text{[Fe/H]}_{\text{CC}} \approx \frac{y_{\text{Ia,CC}} r_{\text{CC}} M_*}{\mu_{\text{Fe}} \Mgeff 10^{\text{[Fe/H]} - 4.5} \ln(10)}
\label{eq:deltafehCC}
\end{equation}

As the equation shows, the longer the time gap between starbursts, the larger the visible metallicity gap would be, as expected. The gap $\Delta$[Fe/H] is sensitive to the underlying effective mixing gas mass $\Mgeff$ and the initial gas metallicity [Fe/H]. The combination of these two criteria shows that such gaps should only be more prominent in metal-poor dwarf galaxies with lower metallicities and effective gas masses.

To illustrate, consider a gas mass of $M_{\text{g}} = 3 \times 10^8\,{\text M}_{\odot}$, a starburst that creates a stellar mass of $M_* = 10^6 \, {\text M}_{\odot}$, and a burst gap of $\delta t_{\text{gap}} = 200$ Myr. At the initial metallicity of [Fe/H] = -2, we find that the resulting $\Delta$[Fe/H]$_{\text{Ia}}$ from Type Ia supernovae in the MDF is approximately 0.15 dex. When combined with the contribution from core-collapse supernovae, this leads to a total metallicity gap of 0.2 dex, which is significantly larger than the typical spectroscopic metallicity precision. Even at [Fe/H] = -1.5, if we assume a $\delta t_{\text{gap}} = 400$ Myr following the simulations, the expected gap is still about 0.11 dex, which should be measurable in high-resolution spectroscopy. The gap is even more pronounced at lower metallicities, as illustrated in Figure~\ref{fig1}. This is the key signal that we aim to detect and quantify in our study.

\begin{figure*}
\centering
\includegraphics[width=1.0\textwidth]{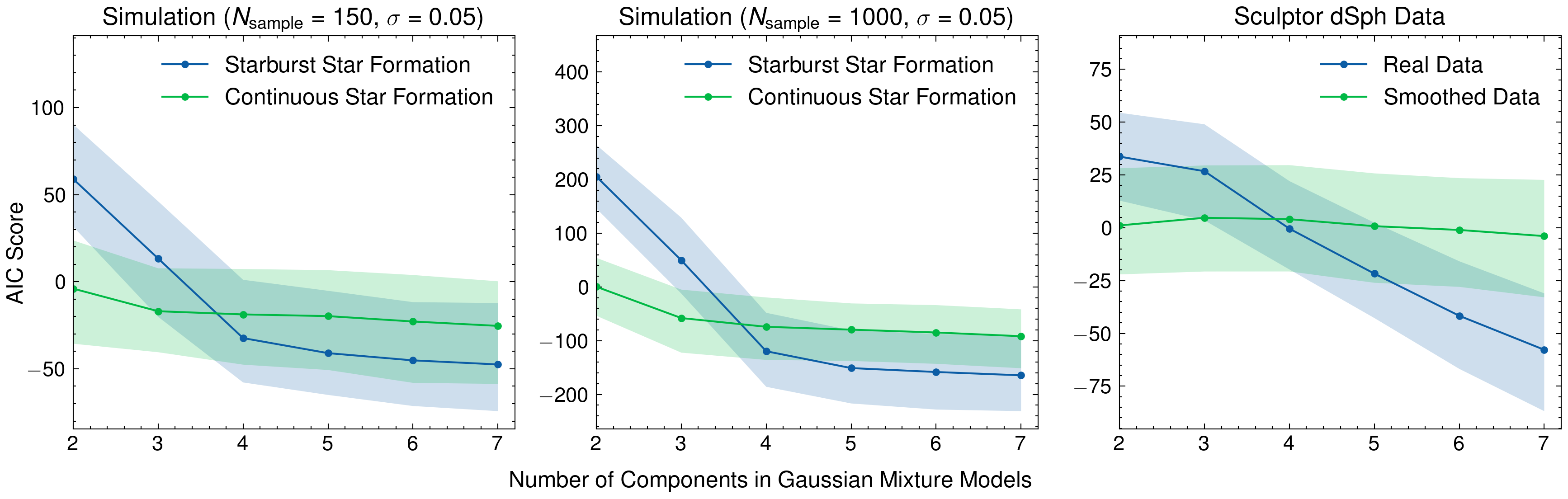}
\caption{Detecting discontinuous star formation by searching for multimodal components in the chemical tracks using Gaussian Mixture Models. The figure shows the Akaike Information Criterion score versus the number of components assumed in the GMM, with a lower score indicating that the model is favored. The shaded region shows the $1\sigma$ uncertainty from bootstrapping. In the bursty star formation scenario, if the measurement precision is comparable to or smaller than the gap between the discontinuous chemical tracks, multiple components would be preferred. The left and middle panels show the results for simulated data, as depicted in Figure~\ref{fig1}, while the right panel presents the case of real APOGEE data for the Sculptor dwarf spheroidal galaxy. The figure demonstrates that, even for a limited sample size of 150 and limited precision in the abundance measurements ($\sigma = 0.05$), there should be a signal in the AIC score despite the uncertainty. We estimate the AIC score for the simulations using 100 realizations and bootstrapping for the real data to estimate the uncertainty. A larger sample of 1000 stars, as shown in the middle panel, will lead to an unambiguous detection of multimodality. In the case of the real data, which more closely aligns with the simulated case in the left panel, we demonstrate statistical evidence for multimodality in the Sculptor dwarf spheroidal galaxy; the AIC score is higher for a larger number of components, superseding the sampling noise, suggestive of bursty star formation. The real data also shows a more prominent signal at larger number of components, presumably due to the fact that, unlike assumed in the simulations, the gas mass varies as a function of time, leading to individual Gaussian modes being more distinct than in the simulations.}
\label{fig2}
\end{figure*}

\section{Measuring the Discontinuous Chemical Track}
\label{sec:measurement-sims}

Our theory developed above suggests that measuring the discontinuous chemical tracks by looking for distinct stripes in the [Mg/Fe]-[Fe/H] plane or gaps in the MDF should be more feasible and provide strong evidence for the bursty nature of star formation in these systems. Given the simplicity of our chemical evolution model, we search for the multi-modal discontinuous track in an unsupervised manner with Gaussian Mixture Models (GMMs) of the [Mg/Fe]-[Fe/H] plane. As the name suggests, GMMs aim to model the chemical space with a mixture of Gaussians, mathematically expressed as:
\begin{equation}
p(\mathbf{x}) = \sum_{k=1}^{K} \pi_k \mathcal{N}(\mathbf{x} | \boldsymbol{\mu}_k, \boldsymbol{\Sigma}_k)
\end{equation}

\noindent
where $\mathbf{x} = (\text{[Fe/H]}, [\text{Mg}/\text{Fe}])$ represents a data point in the chemical space, $K$ is the number of Gaussian components, $\pi_k$ is the mixing coefficient (or weight) of the $k$-th component where $\sum_k \pi_k = 1$, and $\mathcal{N}(\mathbf{x} | \boldsymbol{\mu}_k, \boldsymbol{\Sigma}_k)$ denotes the multivariate Gaussian density with mean $\boldsymbol{\mu}_k$ and covariance matrix $\boldsymbol{\Sigma}_k$.

GMMs are particularly well-suited for our analysis for several reasons. The strong inductive bias - which assumes the data can be modeled as a mixture of Gaussian components - allows us to perform robust modeling with our small dataset ($\sim$100 stars). While such assumption is certainly a limitation, in the absence of measurement uncertainties, chemical evolution tracks follow well-defined one-dimensional paths, with observed spread primarily reflecting measurement uncertainties. And chemical abundance uncertainties are approximately Gaussian, as they derive from combining multiple spectral features. We note that while more flexible density estimation methods exist \citep[e.g., Normalizing Flows][]{Ting2022}, GMMs explicitly model discrete subcomponents - exactly what we expect from episodic star formation. As we will see, to ensure robust results despite these limitations, we carefully validate our GMM fits through bootstrap resampling and comparison with smoothed models (see Figure~\ref{fig2}) with simulated data, finding the method is adequate to find statistical evidence for multimodality in our dataset.

To determine the best-fitting model, we fit GMMs with different numbers of components to the chemical space and compare their likelihoods, defined as:
\begin{equation}
L(\boldsymbol{\theta}) = \prod_{i=1}^{N} p(\mathbf{x}_i | \boldsymbol{\theta})
\end{equation}

\noindent
where $\boldsymbol{\theta} = \{\pi_k, \boldsymbol{\mu}_k, \boldsymbol{\Sigma}_k\}_{k=1}^{K}$ represents the set of all parameters in the GMM, and $N$ is the number of data points. 

Finding the optimal parameters for a GMM is a complex non-convex optimization problem that cannot be solved analytically. We therefore fit GMMs using the Expectation-Maximization (EM) algorithm implemented in scikit-learn, which iteratively estimates the model parameters by alternating between two steps: (1) computing the probability of each data point belonging to each component (E-step), and (2) updating the model parameters to maximize the likelihood given these probabilities (M-step). The EM algorithm is guaranteed to converge to a local optimum, but may not find the global optimum. To address this limitation, we perform 1000 independent fits with different random initializations, validating that each component has at least 4 points and covariance correlations below 0.9, and select the model with the highest likelihood score. This approach helps avoid poor local optima and ensures robust convergence to a well-behaved solution that accurately captures the underlying structure in the data.

It is important to note that GMMs with more components will have a strictly better likelihood than their lower-component counterparts: setting the weight of an additional component to zero to the lower-component case. To properly evaluate the model performance, we need topenalize models based on their complexity. In this study, we use the Akaike Information Criterion (AIC, \citealt{Akaike1974}) defined as:
\begin{equation}
\text{AIC} = 2m - 2\ln(\hat{L})
\end{equation}

\noindent
where $m$ is the number of parameters in the GMM ($K \left(1 + d + (d(d+1))/2 \right) - 1$ for a 2-dimensional GMM, where $K$ is the number of modes and $d=2$ the dimension), and $\hat{L}$ is the maximum likelihood estimate of the model. A lower AIC is better.

Based on the AIC score, we determine the optimal number of Gaussian components that best describe the data. The AIC balances model fit against model complexity according to the equation above. A lower AIC indicates a more favorable trade-off between complexity and fit. Specifically, $-2\ln(\hat{L})$ rewards models that better fit the data, while $2m$ penalizes complex models. Under the assumption that the true model lies within the family of candidate models, AIC is asymptotically optimal in selecting the model closest (in a Kullback--Leibler sense) to the true process \citep{Akaike1974}. When the preferred model includes multiple Gaussian components, this naturally suggests multimodality in the chemical abundance space, pointing to multiple subpopulations. In an astrophysical context, such a pattern is consistent with bursty star formation, wherein different episodes of star formation give rise to distinct chemical loci.

To demonstrate the effectiveness of GMMs in detecting discontinuous chemical tracks, we start with a mock sample as depicted in Figure~\ref{fig1}. We consider two scenarios: the fiducial model with regular starbursts spaced between 150-450 Myr apart (the blue SFH in the top right panel of Figure~\ref{fig1}) with each starburst lasting 150 Myr; and the continuous smooth star formation model that integrates to the same total stellar mass of $10^7 {\rm M}_{\odot}$ (the dashed orange line in Figure~\ref{fig1}). We draw samples according to the respective star formation rates and assume an uncertainty of 0.05 dex in both [Fe/H] and [Mg/Fe], similar to the quoted precision and intrinsic dispersion of APOGEE abundances in Sculptor \citep{Mead2024}.

The results are presented in the left and middle panels of Figure~\ref{fig2}. The left panel shows a realistic case of current observations, a sample size of 150 stars with Mg and Fe measurements \citep{Kirby2011,Hill2019,Abdurrouf2022}. The figure demonstrates that even with current observations, the case drawn from bursty star formation favors is able to resolve at least 4 components, as evident from the continuously decreasing AIC scores with a higher number of components. In contrast, the continuous star formation case exhibits nearly flat AIC scores with respect to the number of components, demonstrating that AIC is a reasonable penalty. However, the shaded bootstrap uncertainties in the AIC values indicate that sampling uncertainty cannot be neglected with only 150 stars.
\begin{figure}
\centering
\includegraphics[width=0.4\textwidth]{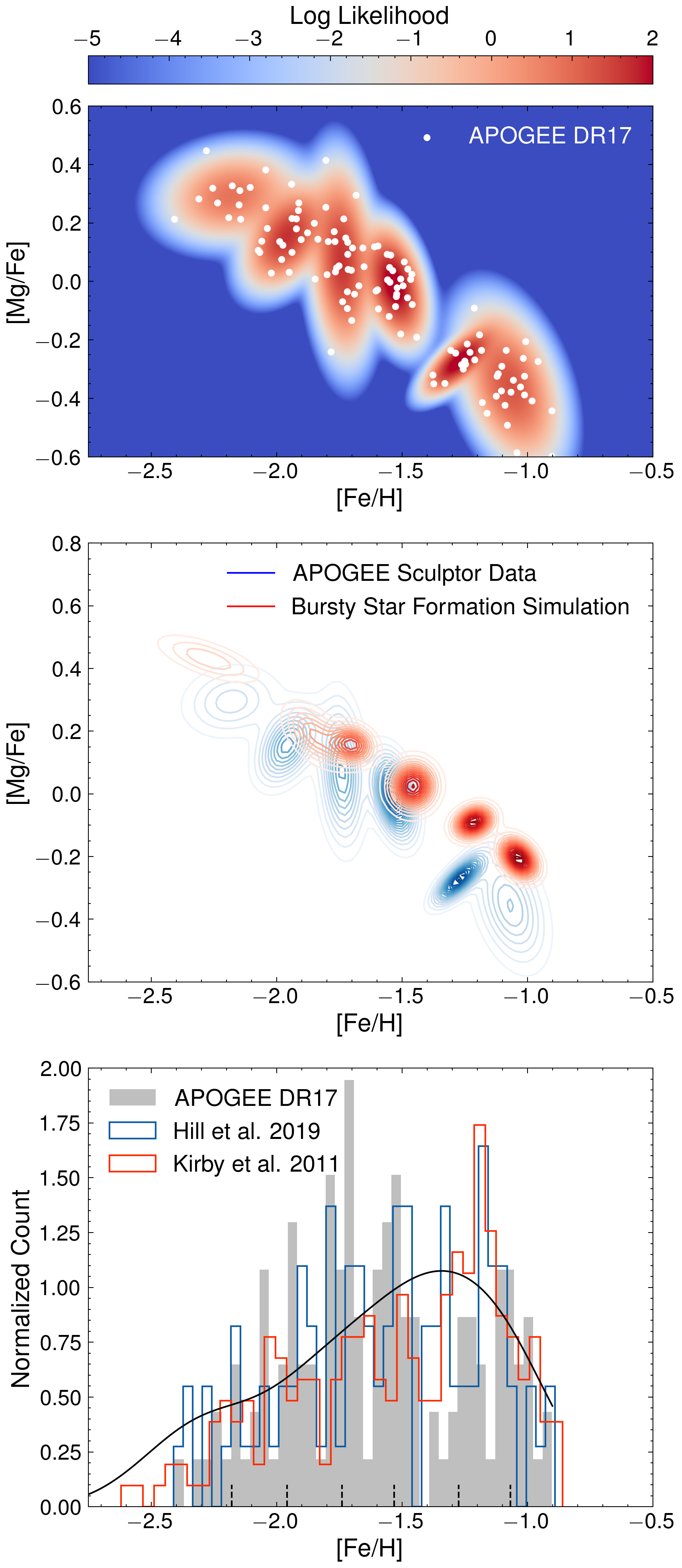}
\caption{The APOGEE Sculptor dwarf spheroidal galaxy data shows multimodality in the [Mg/Fe]-[Fe/H] plane and metallicity distribution function. The top panel displays the APOGEE Sculptor sample, with the background representing the likelihood from the best-fit 6-component Gaussian Mixture Model. The middle panel compares the likelihood contours of the observed data (blue) with a 6-component GMM fit (red) to an equal number of samples drawn from our simulated bursty star formation model. The similarity between the contours of the observed data and the simulated model reinforces the interpretation of episodic star formation in Sculptor. The bottom panel presents the histogram of [Fe/H] for the APOGEE sample, also exhibiting multimodality. The multimodal distribution aligns well with other Sculptor samples from Hill et al. (2019) and Kirby et al. (2013). The vertical lines in the bottom panel indicate the fitted means of the GMM components, and the solid black line shows the smoothed MDF of the simulated sample.}
\label{fig3}
\end{figure}

This situation improves if we can further extend the sample size to 1000 stars, as shown in the middle panel. In this case, the sampling noise on the AIC scores decrease substantially and improve the ability to detect multimodality. Obtaining abundances of over 1000 stars in massive dSphs like Sagittarius, Fornax, and Sculptor will soon be possible with spectroscopic surveys on 4-8m telescopes like 4DWARFS \citep{Skuladottir2023} and Subaru PFS \citep{Takada2014}. The advent of ELTs will enable such measurements in all but the faintest dSph galaxies \citep{Ji2019}. Although not shown, our tests indicate that measurement precision is less critical than sample size for detecting multimodality in these systems. The metallicity gaps, particularly at low [Fe/H], are typically larger than 0.05 dex, exceeding typical spectroscopic uncertainties. However, at higher metallicities ([Fe/H] $> -1.5$) and for shorter starburst gaps ($\sim$100 Myr), higher precision measurements could help better resolve the multimodal structure (see Section~\ref{sec:discussion})

\section{Dectecting Bursty Star Formation in Sculptor dSph}
\label{sec:measurement-sculptor}

We now focus on real observations of the Sculptor (Scl) dSph. The Sculptor dSph is ${\approx}$84 kpc away \citep{MartinezVazquez2015}, with a dynamical mass $\sim 10^8 \, {\rm M}\odot$ \citep{Braddels2012,Salvadori2008} and a present stellar mass of roughly $\sim 3-8 \times 10^6  \, {\rm M}\odot$ \citep{Salvadori2008,deBoer2012,Bettinelli2019} that corresponds to an initially formed stellar mass of $\sim 10^7 \, {\rm M}_\odot$. Sculptor has extensive chemical data \citep{Kirby2011,Hill2019,delosReyes2022,Abdurrouf2022} and star formation histories from color-magnitude diagrams \citep{deBoer2012,Weisz2014,Vincenzo2016,Bettinelli2019} that suggest it formed all its stars in 1-2 Gyr. Sculptor is the ``quintissential'' dSph, and its formation history is likely well-approximated by a well-mixed one-zone chemical evolution model.

We use chemical abundances of Sculptor from APOGEE/ASPCAP DR17 \citep{GarciaPerez2016,Abdurrouf2022}. We selected member stars using proper motion members from \citet{Pace2022} and restricted to stars within three velocity dispersions of the central velocity \citep{Walker2009}. We then restrict to stars with SNR $>30$; good quality flags \texttt{BRIGHT\_NEIGHBOR}, \texttt{VERY\_BRIGHT\_NEIGHBOR}, and \texttt{SUSPECT\_BROAD\_LINES}; good ASPCAP flags \texttt{STAR\_BAD}, \texttt{METALS\_BAD}, and \texttt{ALPHAFE\_BAD}; [Mg/Fe] error $<0.1$ dex; and $\log g < 2$. This results in 124 member stars with high quality abundances.

In the bottom panel of Figure~\ref{fig3}, we show the histogram of metallicity [Fe/H]. Even when projected onto the metallicity distribution, which can smear out the multimodality from starbursts (see the top left panel of Figure~\ref{fig1}), the histogram of the Sculptor dSph shows visually apparent multimodality, with peaks at [Fe/H] = -2.18, -1.96, -1.74, -1.53, -1.27, and -1.07. The peaks are evaluated by fitting a 6-component GMM to the data, and the means of these GMMs are shown as vertical dashed lines at the bottom to guide the eye. For comparison, we overplot the metallicity measurements from \citet{Hill2019} and \citet{Kirby2011}. We shift Kirby's measurements by 0.1 dex in [Fe/H] to match the peak locations seen in the Hill and APOGEE data, which is consistent with a 0.07 dex zero-point offset relative to high-resolution spectroscopy noted in that paper.  Careful examination of the metallicity distributions remarkably shows that the 6 MDF peaks identified in APOGEE have corresponding peaks in the Hill and Kirby data. This pushes the limits of quoted measurement uncertainties by \citet{Kirby2011} and \citet{Hill2019}, but we note that their abundance \textit{precision} is likely better, because their full uncertainties include an estimate of accuracy in addition to precision. The consistency between these three independent datasets lends confidence that the multimodality in Sculptor is real.

In the bottom panel, we also present the smoothed MDF of the simulated bursty chemical evolution model, obtained using a KDE with a smoothing scale of 0.2 dex. The good agreement between the simulated and observed MDF is because we adjusted the fiducial star formation history to match the observations (blue SFH in the top right panel of Figure~\ref{fig1}). Without smoothing, the simulated results would exhibit the same multimodal structure as the observed data, as evident from the contours in the middle panel. 

In the top panel, we show a 6-component GMM fit to the Sculptor data. The colored background in the top panel of Figure~\ref{fig3} shows the likelihood of the best-fit 6-component model, illustrating ing of the data. The middle panel of Figure~\ref{fig3} shows that the contours and locations of the multimodal data are qualitatively consistent with the fiducial simulated models that we constructed. This consistency between the observed data and the simulated bursty star formation models suggests that the Sculptor dwarf spheroidal galaxy has undergone episodic star formation throughout its history.
\begin{figure}
\centering
\includegraphics[width=0.4\textwidth]{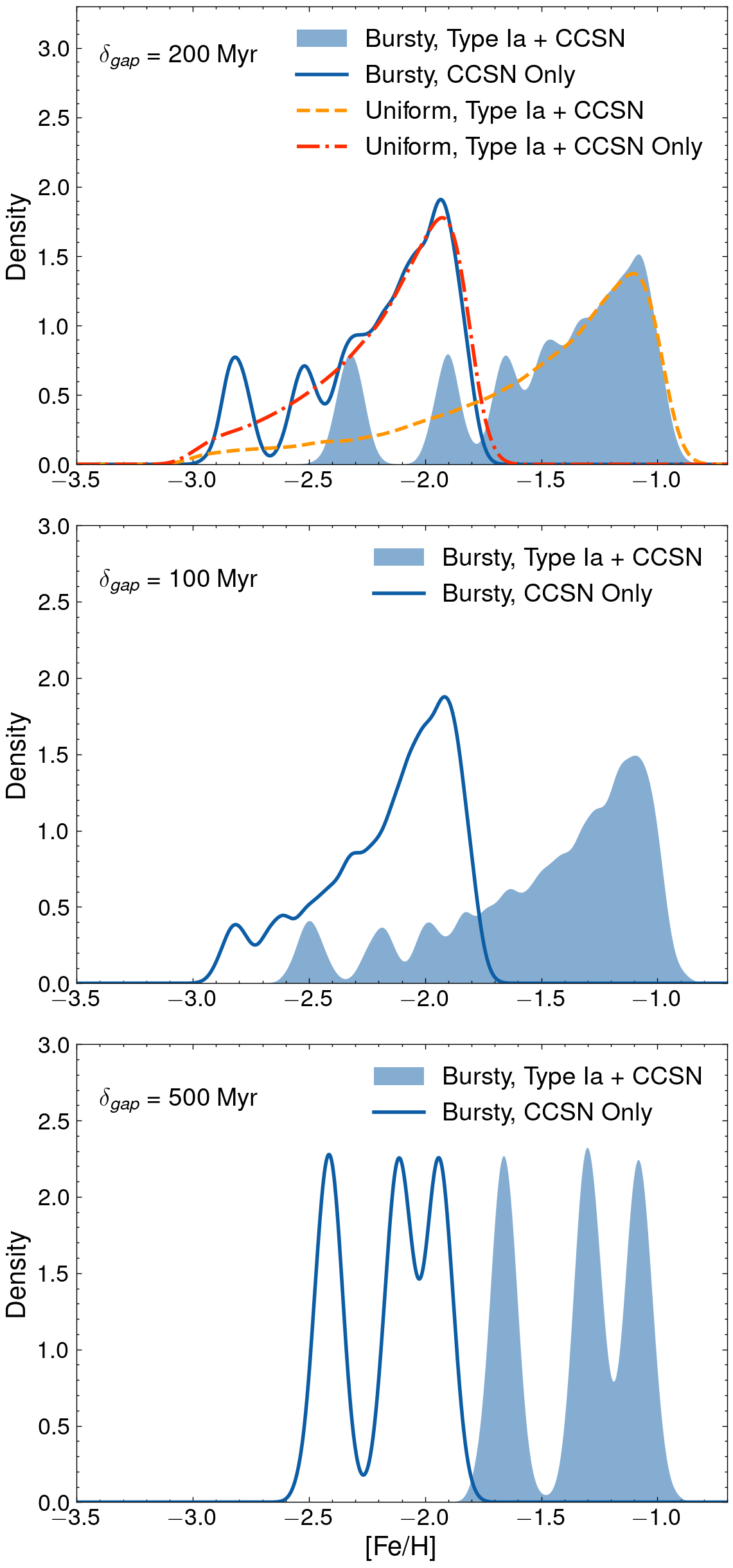}
\caption{The contribution from Type Ia supernovae enhances the prominence of multimodality in the metallicity distribution function. The figure presents the MDF for various star formation histories, each integrating to a total stellar mass of $10^7 {\text{M}}_{\odot}$ over 2 Gyr and assuming a mixing gas mass of $3\times 10^8 {\text{M}}_{\odot}$. The shaded blue region in the top panel depicts the MDF for the fiducial case with instantaneous bursty star formation and a gap ($\delta t_{\text{gap}}$) of 200 Myr, considering contributions from both core-collapse and Type Ia supernovae. The solid blue line represents the case where only core-collapse supernovae contribute, resulting in a smaller and more subdued multimodality in the metallicity distribution. This effect is further suppressed when assuming $\delta t_{\text{gap}} = 100$ Myr, as shown in the middle panel. The bottom panel illustrates the case with $\delta t_{\text{gap}} = 500$ Myr. For comparison, the top panel also includes the MDF for uniform and continuous star formation, represented by the dot-dashed red and dashed orange lines. In these cases, the MDF is smooth and lacks multimodality.}
\label{fig4}
\end{figure}

To more robustly quantify the multimodality, we calculate the AIC score estimate the uncertainty of the AIC score by bootstrapping the data. The AIC score as a function of the number of components is shown in the blue band in the right panel of Figure~\ref{fig2}. Just like in the case of the simulated data in the left panel, the real data also exhibits a similar trend, with the AIC score decreasing as the number of components increases, signaling a multimodal nature in the [Mg/Fe]-[Fe/H] plane. For comparison, we show the AIC score for a smoothed data example as a green band in the right panel. We generate the smoothed data by fitting the real data with a two-component GMM\footnote{We choose two components because there is an apparent kink in the chemical track around [Fe/H] = -1.3 (see top panel of Figure~\ref{fig3}), but the results are practically the same if we were to fit a single Gaussian distribution.}, draw from the distribution, and ensure that the smoothed distribution follows the contours of the data. In this case, the smoothed data shows no trend in the AIC scores, demonstrating that the real data has a statistical tendency to favor the case of multimodality. The close agreement between the AIC behavior of our simulated bursty star formation models and the actual Sculptor data provides statistical evidence linking the observed multimodality to discrete star formation events, particularly given the clear contrast with the flat AIC scores seen in our continuous star formation simulations.

Interestingly, the AIC trend is even stronger for the real data than the simulated data. One potential reason is that in the simulated data we assume a constant effective gas mass. Thus each 2D Gaussian mode in the [Mg/Fe]-[Fe/H] spacehas roughly the same tilt and spread, as shown by the red ellipses in the middle panel of Figure~\ref{fig3}. However, in the real data, as the effective gas mass might vary, which could lead to more distinct Gaussian modes, as shown in the background of the top panel and the blue contours in the middle panel. This would result in a stronger AIC score trend as a function of the number of components, continuing to decrease above four components.

\section{Discussion}
\label{sec:discussion}

In this study, we develop an analytic theory demonstrating that bursty star formation leads to a discontinuous chemical track that is distinguishable from continuous star formation. In particular, we expect to find multiple modes in the [Mg/Fe]-[Fe/H] plane and in the metallicity distribution function. We adopt Gaussian Mixture Models to detect such multimodality, as they are model-agnostic. As a proof of concept, we apply this method to the existing APOGEE data of the Sculptor Sph galaxy. We find that the AIC score trend is distinguishable from the smoothed case. The observed signatures are consistent with other datasets of Sculptor as well as the theory developed, supporting an episodic star formation history of the galaxy.

\subsection{100 Myr Time Resolution for Dwarf Galaxy Star Formation Histories}
\label{subsec:discussion-prominent-signal}

In Figure~\ref{fig4}, we emphasize that the strength of the multimodal signal relies on the production of Fe by Type~Ia supernovae during quiescent periods. The dot-dashed red and dashed orange lines in the top panel show the baseline case where star formation contributes uniformly over 2 Gyr, yielding a total mass of $10^7 \, {\rm M}_{\odot}$ (i.e., dashed orange line in the top-left corner of Figure~\ref{fig1}). In this case, with or without Type Ia supernovae, the MDF is smooththe shaded blue region shows the expected MDF from the starburst case, where we assume instantaneous starbursts of $10^6 {\rm M}_{\odot}$, spanning 2 Gyr with a starburst gap of 200 Myr (i.e., the black line in Figure~\ref{fig1}). In this case, the multimodality is clearly seen at $\mbox{[Fe/H]} \lesssim -1.5$, though it is harder to detect at higher metallicities. If we then remove Type~Ia supernovae, as shown by the solid blue line, the gap is more subdued and becomes invisible after [Fe/H] $> -2.5$. We show a shorter and longer $\delta t_{\text{gap}}$ in the other two panels, illustrating that multimodality can be resolved even down to 100 Myr gaps. 

Such high time resolution in the old star formation histories of local dwarf galaxies is not possible through other means, like color-magnitude diagrams \citep{deBoer2012,Weisz2014,Bettinelli2019}. It is also not resolved by most current chemical evolution models, which impose smooth star formation histories that preclude bursty or episodic star formation \citep[e.g.,][]{Vincenzo2016,Weinberg2017,delosReyes2022,Sandford2024}. 

The episodic star formation we suggest occurs in Sculptor may be consistent with recent JWST observations of high redshift galaxies that display miniquenching \citep{Antwi-Danso2024,Dome2024,Looser2024}, but for galaxies whose stellar mass is much lower than could be detected directly at high redshift. The progenitor of Sculptor has $M_{UV} = -11$ at $z \sim 7$ \citep{BoylanKolchin2015}, which is at least 4 magnitudes fainter than the lowest-mass spectroscopically confirmed galaxies at $z \sim 7$ \citep{Atek2024}. Thus, chemical evolution can provide unprecedented time resolution for the star formation histories of early low mass galaxies.

A natural question is whether even shorter timescales of ${\sim}10$ Myr could be resolved by chemical evolution. Indeed we assumed that all CCSN produce Mg at a single time, but in reality there is a delay time distribution for CCSNe as well lasting from 3-30\,Myr \citep[e.g.,][]{delosReyes2022}. We found this is unlikely to be significant for most galaxies, as repeated bursts on this timescale quickly wash out any potential signatures. However, the faintest dwarf galaxies forming from a single feedback-quenched starburst \citep[e.g.,][]{Gallart2021,Chen2022} may still exhibit these signatures.

\subsection{Intrinsic spread in [$\alpha$/Fe] from the starbursts}
\label{subsec:mgfe-spread}

Initially, we had anticipated that the [Mg/Fe] spread would be an independent useful indicator of bursty star formation. This was motivated by past studies \citep[e.g.,][]{Recchi2015,Muley2021,Patel2022,Mead2024} and by the sharp sawtooth [Mg/Fe] vs [Fe/H] pattern in Figure~\ref{fig1}. However, detecting and using the [Mg/Fe] spread turns out to be substantially more difficult than the [Fe/H] gaps in the model presented here. Observationally, this is because it is much harder to detect an intrinsic scatter than a difference in means: the uncertainty on intrinsic scatter scales as $N_*^{1/4}$ while means scales as $N_*^{1/2}$ \citep[see][]{Ting2022}. Theoretically, it is because star formation bursts have to be somewhat fine-tuned to capture the maximum [Mg/Fe] spread: a burst must be longer than 10 Myr to capture any Mg being produced, and shorter than 100 Myr so Type~Ia supernovae do not start contributing to Fe. Even when this happens, the signal is quite small.

Following Section~\ref{sec:motivation}, we can approximate the expected spread in [Mg/Fe] with:
\begin{equation}
\begin{alignedat}{2}
\Delta\text{[Mg/Fe]} \approx & \frac{y_{\text{CC,Mg}} r_{\text{CC}} M_*}{\mu_{\text{Mg}} \Mgeff 10^{\text{[Mg/H]}-4.4} \ln(10)} \\
& \qquad - \frac{y_{\text{CC,Fe}} r_{\text{CC}} M_*}{\mu_{\text{Fe}} \Mgeff 10^{\text{[Fe/H]}-4.5} \ln(10)}
\end{alignedat}
\label{eq:deltamgfe}
\end{equation}
where we have neglected the SNIa contribution assuming that it is small in the burst.\footnote{The exponent for [Mg/H] has the term -4.4 instead of -4.5 (as seen in [Fe/H]) due to a slight difference in the solar normalization values for magnesium and iron. According to \citet{Asplund2009}, there's a 0.1 discrepancy between these elements.} Assuming [Mg/Fe] = 0.4, an integrated star formation of $M_* = 10^6 \, {\text M}_{\odot}$, and a gas mass of $\Mgeff = 3 \times 10^8 \, {\text M}_{\odot}$, we estimate that $\Delta$[Mg/Fe] changes by only 0.08 dex at [Fe/H] = -2, decreasing to 0.02 dex at [Fe/H] = -1.5. However, our assumption of instantaneous mixing into a large constant effective gas mass for CCSN may hinder our ability to resolve this signal, and concentrating the Mg into a smaller gas mass would enhance the signal \citep[e.g.,][]{Emerick2018,Patel2022}.

\subsection{Quantitative Interpretation of the [Fe/H] Gaps in Sculptor}
\label{subsec:discussion-sculptor}

In Section~\ref{sec:measurement-sculptor} and the bottom panel of Figure~\ref{fig3}, we see that the spacing between adjacent bursts varies from 0.20 to 0.25 dex, with larger separations occurring at lower metallicities, which is quite uniform. From Equation~\ref{eq:deltafeh}, assuming Type Ia supernovae contribution dominates the gap, this implies that the quantity $(M_*/\Mgeff) [(t_{\text{min}}^{-0.1}- t_{\text{gap}}^{-0.1})/10^{\text{[Fe/H]}}]$ remains approximately constant throughout the galaxy's evolution. In our model, we have assumed $(M_*/\Mgeff)$ is constant, and adjusted $t_{\text{gap}}$ to match this observed spacing. As shown in the middle panel of Figure 3, we find that a gradual increase of $\delta t_{\text{gap}}$ from 150 Myr to 450 Myr with an increment of 50 Myr per starburst (see the top right panel of Figure~\ref{fig1}) leads to a good qualitative match to the data (see the middle panel of Figure~\ref{fig3}). It was not required that the burst separations fit into the 2 Gyr duration of Sculptor's formation history, but it does, suggesting that taking $\Mgeff$ to be constant is a reasonable approximation.

Furthermore, we can try to interpret the intrinsic scatter in [Mg/Fe]. We estimate the intrinsic scatter of each component from our GMM fit, taking the fitted covariance matrices $\Sigma_k$ and subtracting 0.05 dex in quadrature as a typical uncertainty. The measured scatters for the three prominent modes at [Fe/H] $\simeq$ -1.74, -1.53, and -1.27 are found to decrease from 0.13, 0.08, to 0.05 dex, respectively. Following Equation~\ref{eq:deltamgfe}, if we take $\Mgeff$ to be constant, the sequence of Mg scatter closely follows the ratio $10^{\text{[Fe/H]}}$, i.e. that the scatter can be completely explained by enrichment into a constant effective gas mass \citep[also see][for a similar conclusion on $r$-process scatter in the halo]{Brauer2021}.

These calculations are clearly primitive, and a full interpretation should more carefully consider evolution of $\Mgeff$. However, the fact that we could find a suitable bursty star formation history for Sculptor that matches the observed CMD-based star formation history, metallicity distribution function, and [Mg/Fe] vs [Fe/H] evolution without needing $\Mgeff$ evolution suggests that it should not vary too much.

\subsection{Starburst Signals from Other Dwarf Galaxies}
\label{subsec:discussion-other-galaxies}

While we chose to study the Sculptor dSph galaxy in this work due to the availability of multiple datasets that corroborate our results and its status as one of the most well-studied dwarf galaxies, other galaxies could potentially show evidence for bursty star formation.

\begin{figure*}
\centering
\includegraphics[width=0.8\textwidth]{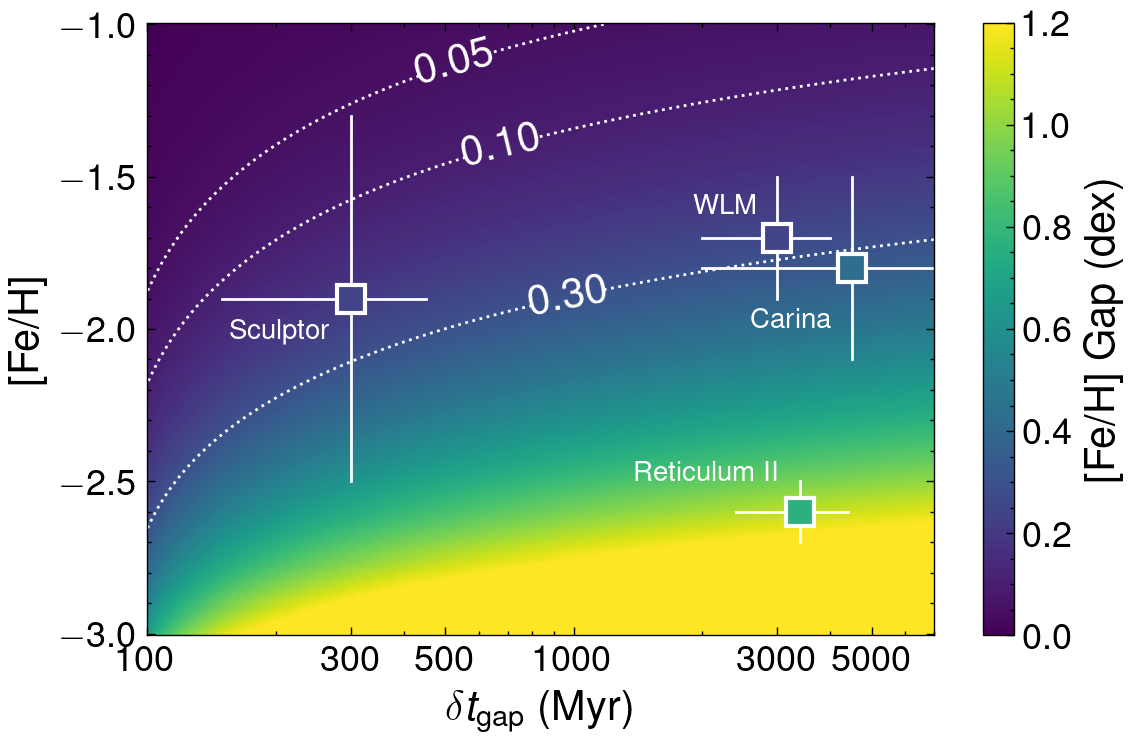}
\caption{The predicted metallicity gap in the metallicity distribution function as a function of initial metallicity and starburst time gap. The white dotted lines include the typical uncertainties of metallicity measurements from high-resolution spectroscopy ($R \simeq 20,000$, 0.05 dex), mid-resolution spectroscopy ($R \simeq 6,000$, 0.1 dex), and photometric metallicities (0.3 dex). Overplotted are measurements from a few dwarf galaxies. The Sculptor dwarf galaxy data points are measured in this study by matching the [Mg/Fe]-[Fe/H] contours. For the other galaxies, Carina, WLM and Reticulum II, the starburst time gap is estimated from color-magnitude diagrams, and the metallicity gap (represented by the color of the symbols) is determined from the [Fe/H] distribution. Despite the simplicity of our analytic model, it accurately predicts the metallicity gaps when assuming an integrated SFR to mixing gas mass ratio of 1/30.}
\label{fig5}
\end{figure*}

To demonstrate the expected signal, in Figure~\ref{fig5}, we show the expected metallicity gap as a function of the time gap in star formation, $\delta t_{\text{gap}}$, as well as the metallicity of the galaxy, following the theory developed in Section~\ref{subsec:feh-gap}\footnote{To be more exact, we choose not to use the Taylor expansion (Equation \ref{eq:taylor-expansion}), but rather perform the full numerical evaluation for this calculation.}. The solid white lines in the figure represent $\Delta$[Fe/H] values of 0.05 dex, 0.1 dex, and 0.3 dex, which signify the typical uncertainty of metallicity from high-resolution ($R \simeq 20,000$), mid-resolution ($R \simeq 6,000$), and photometric surveys of stars, respectively. As the predicted metallicity gap depends on the ratio of the integrated star formation history to the mixing gas mass, here, we assume a total integrated star formation history that leads to $10^7 {\text M}_{\odot}$ and a mixing gas mass of $3\times 10^8 {\text M}_{\odot}$, i.e., a ratio of 1/30.

Very few dwarf galaxies have enough metallicities to directly measure a multimodal MDF on the 100s of Myr timescales that we propose for Sculptor. However, many galaxies have clear stagnant periods lasting ${>}1$ Gyr that are clearly distinguished using main sequence turnoffs in color-magnitude diagrams (CMDs). Table~\ref{tab:tgap} presents a compilation of several such galaxies, showing the time gap between significant star formation bursts for WLM \citep[$M_\star \sim 10^8 {\rm M}_\odot$][]{McQuinn2024}, Carina ($M_\star \sim 10^6 {\rm M}_\odot$, \citealt{Norris2017a}, which has three distinct bursts), and Reticulum II \citep[$M_\star \sim 10^4 {\rm M}_\odot$][]{Simon2023}. We then estimate the sizes of [Fe/H] gaps using jumps in the metallicity distribution functions occurring at similar cumulative fractions as the star formation histories \citep{Leaman2013,Norris2017b,Ji2023} or a jump in the age-metallicity relation \citep{McQuinn2024}. In practice, this is usually a gap between the most metal-poor component and the bulk of star formation. The exception is Reticulum~II, where the metallicity distribution has a secondary metal-rich peak at $\mbox{[Fe/H]}=-2.0$ (Luna et al., in prep). The CMD-based histories only resolve Gyr-length gaps, but our formalism should still predict the size of the gap due to continued Fe production by SN~Ia during the quiescent period. 

The predicted [Fe/H] gaps generally align well with the observed data, with Reticulum II being a potential exception. This discrepancy can be resolved if we adjust the ratio of mixing gas mass to stellar mass, which is expected due to higher mass loading factors in lower mass galaxies \citep{Alexander2023,Sandford2024}. We found that increasing this ratio from 30 to 70 would bring the predicted metallicity gap for Reticulum II into agreement with observations. 
Our initial investigation makes very simple assumptions about the bursty star formation history and detailed gas mass history, and we plan to address these assumptions in future work. However, the very fact that this simple prescription already reasonably reproduces the age and [Fe/H] gaps in dwarf galaxies suggests that the basic setup is correct.

\begin{table*}[]
    \centering
    \setlength{\tabcolsep}{10pt} 
    \begin{tabular}{ccccccc}
    \hline
         Galaxy & $\delta t_{\rm gap}$ (Gyr) & [Fe/H] & Observed [Fe/H] Gap & Predicted [Fe/H] Gap* & References \\ \hline
         Sculptor & $0.3 \pm 0.15$ & $-1.9 \pm 0.6$ & $0.20-0.26$ & 0.20 & This Work \\
         WLM & $3 \pm 1$ & $-1.7 \pm 0.2$ & 0.3 & 0.26 & 1 \\
         Carina & $4.5 \pm 2.5$ & $-1.8 \pm 0.3$ & 0.35 & 0.36 & 2,3 \\
         Reticulum II & $3.4 \pm 1$ & $-2.6 \pm 0.1$ & 0.64 & 1.09 (0.65*) & 4,5 \\
         \hline
    \end{tabular}
    \caption{Comparison of star formation quiescent periods ($\delta t_{\rm gap}$) and metallicity gaps in dwarf galaxies. For each galaxy, we list the measured time gap between star formation episodes, the metallicity at which the gap is measured, the observed metallicity gap in the [Fe/H] distribution, and our model's predicted gap size.  \\ $\quad$\\ References. 
    1: \citealt{McQuinn2024},
    2: \citealt{Norris2017a},
    3: \citealt{Norris2017b},
    4: \citealt{Simon2023},
    5: \citealt{Ji2023} and Luna et al. in prep.\\
    *: The predicted [Fe/H] gap is based on a stellar mass to mixing gas mass ratio of 1/30. The value in brackets shows the prediction if we assume a ratio of 1/70 instead.
    }

    \label{tab:tgap}
\end{table*}

As shown in Figure~\ref{fig5}, a typical mid-resolution survey with a relative precision of metallicity of 0.10 should be able to detect a metallicity gap at a relatively low metallicity of [Fe/H] = -2, as the metallicity gap in the MDF is predicted to be larger than the uncertainty, even with a time gap of only $\delta t_{\text{gap}} = 150$ Myr. A high-resolution survey can extend this limit to [Fe/H] = -1.7. The former is consistent with what we see in the Sculptor dSph. As shown in the bottom panel of Figure~\ref{fig3}, while the sample from \citep{Kirby2011}, which is from mid-resolution spectra, has a smaller precision than APOGEE and \citep{Hill2019}, the metallicity gap in the MDF is still detectable, albeit more smoothed out. For quiescent periods longer than 0.5 Gyr, mid-resolution surveys should detect this gap even for a higher metallicity of [Fe/H] = -1.6. Even with the somewhat poor precision of photometric metallicity, assumed to be 0.3 dex, it remains possible to detect the metallicity gap for quiescent periods longer than 0.5 Gyr and with a metallicity of about -2.2.

\subsection{Caveats and Future Directions}
\label{subsec:discussion-caveats}

A key strength of studying episodic star formation through gaps in elemental abundances, rather than relying solely on color-magnitude diagrams, is the improved temporal resolution. However, this resolution is still largely constrained by the delay time of Type Ia supernovae. It is crucial to recognize that the actual star formation history is likely more complex, potentially comprising numerous mini-scale bursts on timescales of 10-30 Myr within the `starbursts' identified in this study \citep[e.g.,][]{Wheeler2019,Patel2022}. Our approach thus provides a high-level view of the episodic star formation history, focusing on major starbursts defined by sufficiently long quiescent periods.

Our detection hinges on the fact that, statistically, the Gaussian Mixture Models prefer a higher number of components. The advantage of this method is that it aims to detect multimodality without requiring a detailed characterization of the measurement uncertainty of the abundances. However, the implicit assumption we made here is that the measurement uncertainty in the elemental abundances is unimodal and smoothly varying across the [Mg/Fe]-[Fe/H] plane. We consider this to be a fairly minimal assumption, but there are some pathological effects that might cause artificial clustering. For example, our detected MDF peaks in APOGEE are spaced between 0.20 and 0.26 dex apart, which is close to the 0.25 dex grid spacing of the ASPCAP grid \citep[][and online DR17 documentaion]{Jonsson2020} and could indicate ``noding'' in the interpolation scheme \citep[e.g.,][]{Koposov2024}. However, the peak locations coincide with results from \citet{Kirby2011} (using a different model atmosphere grid and spacing) and \citet{Hill2019} (using equivalent widths), lending some confidence that these are real features.

In this study, we use a constant effective mixing gas mass $\Mgeff$. While this simplifies the description of the metallicity gap, it does so at the cost of making the gas mass less interpretable: $\Mgeff$ is a degenerate combination of gas and metal inflows and outflows. In contrast, standard chemical evolution models tie the gas mass (usually the cold gas within the galaxy) to the star formation and outflow rates through a star formation efficiency and mass loading factor or prescription \citep[e.g.,][]{Cescutti2008,Salvadori2008,Ishimaru2015,Weinberg2017,Johnson2023}. While such parameterizations are physically well-motivated and provide good results for overall dwarf galaxy scaling relations, current implementations prohibit the type of bursty star formation seen in hydrodynamic simulations.

The effectiveness of our constant $\Mgeff$ assumption likely varies with galaxy mass. More massive dwarf galaxies with deeper potential wells may maintain more stable gas reservoirs, making our assumption more applicable. Conversely, ultra-faint dwarfs with stronger feedback effects might require more sophisticated gas mass evolution models. However, our successful reproduction of Sculptor's chemical patterns suggests that for intermediate-mass dwarf spheroidals, the constant $\Mgeff$ approximation captures the essential physics needed to interpret chemical discontinuities as markers of bursty star formation. Future work incorporating time-varying gas masses will help quantify how feedback efficiency impacts the visibility and interpretation of these chemical signatures across different mass scales \citep[e.g.,][]{Agertz2013,Read2016}.

\section{Conclusion}
\label{sec:conclusion}

In this initial foray into the impact of episodic star formation on chemical evolution, we develop a model to describe bursty star formation and its impact on the chemical properties of stars. We have demonstrated that episodic star formation in dwarf galaxies leads to distinct chemical signatures compared to continuous star formation scenarios. The multimodal chemical signature provides access to timescales much shorter than possible from color-magnitude diagram analysis, particularly for metal-poor systems. Our key findings are:

\begin{itemize}
\item Bursty star formation leads to distinct metallicity gaps and multimodality in the [$\alpha$/Fe]-[Fe/H] space, with the gap primarily contributed by Type Ia supernovae.

\item These signals are prominent and should be easily detectable even with mid-resolution spectroscopic surveys. For systems with metallicity around -2, even a quiescent period of 200 Myr would lead to a gap that can be distinguished with a measurement precision of 0.1 dex.

\item Chemical signatures of bursty star formation are already detectable, with sufficient statistical significance according to the Akaike Information Criterion, in current datasets of $\mathcal{O}$(100) stars. However, larger samples remain important for reducing sampling noise and improving the robustness of the results.
\end{itemize}

As a proof of concept, we apply our model to the APOGEE data of the Sculptor dSph galaxy. Our findings are:

\begin{itemize}
\item Using Gaussian Mixture Models and the Akaike Information Criterion, we show that the [Mg/Fe]-[Fe/H] space of Sculptor favors multiple components, signaling bursty star formation. 

\item We reproduce these results through simulations, showing that they largely match the observed characteristics. The results are further confirmed with independent datasets from \citet{Hill2019} and \citet{Kirby2011}.

\item Based on the models, we find that Sculptor has undergone multiple bursts with quiescent periods ranging from 150-450 Myr within its short 2 Gyr star formation history. This aligns with current JWST studies of miniquenching in high redshift dwarf galaxies, but extending those conclusions 4 magnitudes fainter than can currently be directly seen with JWST.

\item We find that the same simple theory also extends to other dwarf galaxies with Gyr-long quiescent periods that could be probed with color-magnitude diagrams, including WLM, Carina, and Reticulum II. 
\end{itemize}

While our theory and analysis are admittedly simplistic, this study paves the way for a more comprehensive understanding of the complex interplay between episodic star formation, feedback, and the mixing gas mass in dwarf galaxies. Crucially, our work provides a theoretical foundation to interpret the wealth of data that will be obtained in upcoming large spectroscopic surveys and the era of Extremely Large Telescopes, where precise chemical abundances can be obtained for $\mathcal{O}(1000)$ stars. Previous models enforcing smooth chemical evolution histories do not actually require such large numbers of precise abundances, as only a few stars spanning the metallicity range are needed to define a mean chemical evolution trend. However, the bursty models proposed here actually do require large numbers of chemical abundances with high precision. We encourage the development of these and other models that take full advantage of the recent and upcoming major advances in chemical abundance quality and quantity.

\vspace{0.5cm}

We gratefully thank the Research School of Astronomy and Astrophysics at the Australian National University for supporting A.P.J.'s visit as part of the Distinguished Visitor Program, during which this work was initiated. We also appreciate the serene environment of Canberra, the company of marsupials, and the abundant rabbit population on the ANU campus. The environment provided an ideal setting for in-depth discussions about this project. The inspiration for this study arose from a conversation with Andrew Wetzel at UC Davis during Y.S.T.'s recent visit. We also thank Rohan Naidu, Pratik Gandhi, Erwin Chen, Yanjun Sheng, Guilherme Limberg, Evan Kirby, Daniel Weisz, Kim Venn, Andrey Kratsov, and David Weinberg for insightful discussions over the years. Y.S.T.'s research is supported by the ARC DECRA Research Award DE220101520. A.P.J. is supported by the US National Science Foundation grants AST-2206264 and AST-2307599.


\bibliography{manuscriptNotes.bib}
\bibliographystyle{aasjournal}

\end{CJK*}
\end{document}